\documentclass[aip,jcp,amsmath,amssymb,reprint]{revtex4-1}

\usepackage{graphicx}
\usepackage{dcolumn}
\usepackage{bm}
\usepackage{etoolbox}
\usepackage{color}

\newcommand{\p}{\mathrm{P}}
\renewcommand{\i}{\mathrm{I}}
\newcommand{\s}{\mathrm{S}}
\newcommand{\f}{\mathrm{F}}

\newcommand{\sinc}{\mathrm{sinc}}
\newcommand{\e}{\mathrm{e}}
\newcommand{\vac}{\text{vac}}
\newcommand{\twin}{\text{twin}}

\newcommand{\SE}{\mathrm{SE}}

\newcommand{\rephasing}{\mathrm{(r)}}
\newcommand{\nonrephasing}{\mathrm{(nr)}}

\newcommand{\adagger}{\hat{a}^\dagger}
\renewcommand{\a}{\hat{a}}
\newcommand{\tr}{\mathrm{tr}}

\renewcommand{\Re}{\mathrm{Re}}

\newcommand{\E}{\hat{E}}

\makeatletter
\def\@email#1#2{%
 \endgroup
 \patchcmd{\titleblock@produce}
  {\frontmatter@RRAPformat}
  {\frontmatter@RRAPformat{\produce@RRAP{*#1\href{mailto:#2}{#2}}}\frontmatter@RRAPformat}
  {}{}
}%
\makeatother
\begin{document}


\title{Two-dimensional fluorescence spectroscopy with quantum entangled photons: Idler-referenced timing without pump detection}
\author{Yuta Fujihashi}
 \email{fujihashi@g.ecc.u-tokyo.ac.jp}
 \affiliation{Department of Chemistry, The University of Tokyo, Tokyo 113-0033, Japan}
\author{Akihito Ishizaki}%
 \email{ishizaki@chem.s.u-tokyo.ac.jp}
 \affiliation{Department of Chemistry, The University of Tokyo, Tokyo 113-0033, Japan}


\begin{abstract}
Entangled photons have attracted increasing interest as resources for developing time-resolved spectroscopic techniques. Theoretical studies suggest that their non-classical correlations enable time-resolved spectroscopy with monochromatic pumping and can selectively isolate specific Liouville pathways in nonlinear optical signals. In an earlier study, we proposed a fluorescence detection scheme that could, in principle, be implemented using existing single-photon detectors [Y. Fujihashi {\it et al.}, Sci. Adv. {\bf 12}, eaed7026 (2026)]. In that design, the time origin was defined by detecting the arrival of the pulsed laser used to pump the nonlinear crystal for spontaneous parametric down-conversion, a requirement that made the overall experiment cumbersome. This study theoretically examines an alternative protocol that defines the reference time based on the arrival of idler photons. We demonstrate that this idler-referenced scheme functions effectively when the entangled photons exhibit either negative or negligible frequency correlations. Eliminating the pump-timing channel simplifies the optical layout and lowers the experimental barrier to realizing time-resolved two-dimensional fluorescence spectroscopy with entangled photons. Although the photons may exhibit frequency correlations in isolation, their frequency-time degrees of freedom can behave as effectively uncorrelated when considered over the full measurement timescale. Therefore, fully exploiting non-classical correlations requires an entangled photon source whose temporal characteristics are carefully matched to the overall timescale of the experiment.
\end{abstract}

\maketitle


\section{Introduction}
Quantum light is being actively investigated as a resource for optical spectroscopy, offering higher sensitivity than classical-light methods and potentially enabling selective excitation. \cite{Schlawin:2018ci,Eshun:2022en} Following the demonstrated success of linear absorption spectroscopy using quantum entangled photons, \cite{Scarcelli:2003re,Yabushita:2004hy,Kalashnikov:2016cl,Matsuzaki:2022su,Mukai:2022qu,Tashima:2024ul,Matsuzaki:2024su} attention has turned to their application in nonlinear spectroscopic measurements. \cite{Roslyak:2009cy,Raymer:2013kj,Dorfman:2014bn,Schlawin:2016er,Dorfman:2016da,Zhang:2022en,Dambal:2025qu} Two principal strategies have been proposed for nonlinear spectroscopy that use only entangled photons. First, both photons interact with the molecular sample. \cite{Ishizaki:2020jl,Asban:2021di,Chen:2022mo,Gu:2023ph,Fujihashi:2023pr,Kizmann:2023qu,Yadalam:2023qu,Fujihashi:2024pa,Kim2024:pr,Fan:2024en,Jadoun:2025pa,Hong:2025fe} Second, one photon excites the sample, and a coincidence measurement is performed between the resulting fluorescence and its partner photon. \cite{Fujihashi:2020ep,Li:2023si,Harper:2023en,Eshun:2023fl,Gabler:2025be,Alvarez:2025co,Li:2025co,Tsao:2025he} In the first approach, non-classical correlations between photons enable time-resolved spectroscopy with monochromatic pumping \cite{Ishizaki:2020jl,Chen:2022mo,Gu:2023ph,Fujihashi:2023pr} and provide Liouville pathway selectivity that suppresses specific nonlinear contributions. \cite{Asban:2021di,Kizmann:2023qu,Yadalam:2023qu,Jadoun:2025pa} Moreover, theoretical studies have demonstrated that the coincidence detection of an entangled pair can isolate the stimulated emission pathway, producing time-resolved spectra that are substantially clearer than those obtained with classical light sources. \cite{Fujihashi:2024pa,Kim2024:pr,Hong:2025fe} However, the nonlinear optical signal generated when both photons illuminate the sample is extremely weak, making this first strategy experimentally challenging with current photon detection technology. \cite{Landes:2021ex,Parzuchowski:2021se,Raymer:2021la}
The use of squeezed light has been explored as one possible route to overcoming this limitation. \cite{Dorfman:2021mu,Eto:2021en,Cutipa:2022br,Panahiyan:2022tw,Fujihashi:2023pr,Hashimoto:2024fo,Eto:2025sp,Zhang:2025qu,Jadoun:2025mu,Jadoun:2025tw,Fan:2025qu,Schlawin:2025th}

Owing to such practical constraints, interest has increasingly shifted to the second strategy, which applies entangled photons to time-resolved fluorescence measurements. In the second strategy, the spectroscopic information provided by the frequency-time correlations between entangled photons can, in principle, be obtained with classical light; \cite{Albarelli:2023fu,Ko:2023em} however, entangled photons are still expected to offer practical advantages, particularly through a simpler experimental setup. Scarcelli {\it et al.} first proposed recording steady-state fluorescence spectra through the coincidence detection of entangled pairs. \cite{Scarcelli:2008en} Subsequently, a theoretical framework was designed for time-resolved fluorescence based on the same principles. \cite{Fujihashi:2020ep} 
In this approach, one photon of the entangled pair excites the molecular system, while the other serves as a reference in coincidence detection. This configuration allows fluorescence events associated with entangled photon pairs to be selectively extracted while suppressing background contributions from uncorrelated photons, thereby making experimentally detectable spectroscopic signals feasible. Indeed, several experiments have been conducted on fluorescence lifetime using entangled photons on organic molecules, \cite{Harper:2023en,Eshun:2023fl,Gabler:2025be} photosynthetic proteins, \cite{Li:2023si,Li:2025co} and quantum dot solids. \cite{Tsao:2025he} In a previous study, \cite{Fujihashi:2025tw} we extended the existing theory \cite{Fujihashi:2020ep} to include frequency resolution and theoretically demonstrated that time-resolved two-dimensional fluorescence spectra can be observed with entangled photons using current technology, involving a delay-line anode detector combined with a spectrometer. \cite{Iso:2025ca,Iso:2025hy} The approach exploits non-classical correlations between photon pairs to obtain information on the two-dimensional fluorescence spectrum, eliminating the need for multiple pulsed lasers. {\'A}lvarez-Mendoza, {\it et al.} \cite{Alvarez:2025co} experimentally validated a scheme similar to that proposed in an existing study. \cite{Fujihashi:2025tw} By combining a translating wedge-based identical-pulse-encoding interferometer with single-photon detectors, non-classical correlations between entangled photon pairs enabled the recording of time-resolved fluorescence spectra with monochromatic pumping.

\begin{figure}
    \centering
    \includegraphics{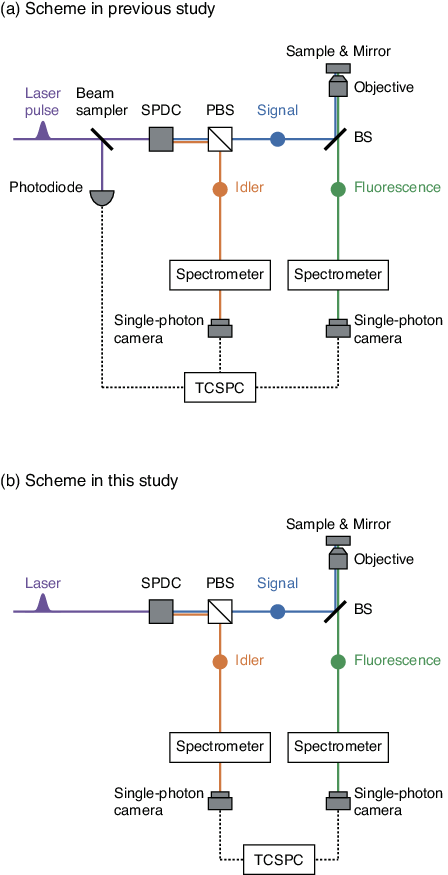}
    \caption{(a) Schematic of the quantum spectroscopy configuration proposed in a previous study \cite{Fujihashi:2025tw}. (b) Setup considered in this study. BS: beam splitter; SPDC: spontaneous parametric down-conversion; PBS: polarized beam splitter; TCSPC: time-correlated single-photon counting device.
    }
    \label{fig1}
\end{figure}

As shown in Fig.~\ref{fig1}(a), the existing scheme \cite{Fujihashi:2025tw} defines time to zero by diverting a fraction of the pulsed laser that pumps the nonlinear crystal for spontaneous parametric down conversion using a beam sampler and detecting the arrival of the pulse. This requirement complicates the detection hardware and necessitates operating the pump laser in pulsed mode, which together make the experiment cumbersome. By contrast, previous fluorescence-lifetime and time-resolved fluorescence studies with entangled photons defined the time origin simply by the arrival of the idler photon, without monitoring the pump. \cite{Harper:2023en,Eshun:2023fl,Gabler:2025be,Li:2023si,Tsao:2025he} Extending this idler-triggered timing to two-dimensional fluorescence spectroscopy can significantly simplify the originally proposed measurement system.

In this study, we assess the feasibility of an idler-triggered scheme for time-resolved two-dimensional fluorescence spectroscopy. We examine the influence of frequency correlations between the entangled photons and the relevant time scales on both the measurement protocol and the molecular information that can be retrieved. Further, we demonstrate that the idler-triggered scheme is feasible when the entangled photons exhibit either negative or negligible frequency correlations.

\section{Optical Configuration}
Before providing a detailed theoretical description, we briefly outline the difference between the scheme proposed in this study and that reported in existing research. \cite{Fujihashi:2025tw}

\begin{figure}
    \centering
    \includegraphics{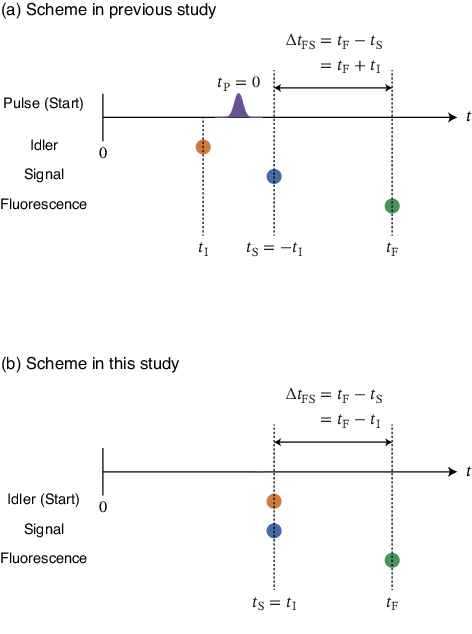}
    \caption{Schematic of detection schemes corresponding to the two experimental configurations, as shown in Fig.~\ref{fig1}. (a) In the pump-referenced setup, the detector system records the arrival-time difference between the photodiode and each single-photon camera. Setting the pump-pulse center to $t_\p=0$ and considering entangled pairs with the negative temporal correlation $t_\s=- t_\i$, the elapsed time from excitation by the signal photon to fluorescence emission can be expressed $\Delta t_\mathrm{FS}=t_\f-t_\s=t_\f+t_\i$. Accordingly, plotting the coincidence counts as a function of $t_\f+t_\i$ yields a time-resolved picture of the excited-state dynamics of the molecular sample. (b) In the idler-referenced scheme, the detector system records only the arrival-time difference between the two single-photon cameras, $t_\f-t_\i$. The figure illustrates the case of negatively frequency-correlated photon pairs, which are generated almost simultaneously within the time window of the entanglement time $T_\e$. Under this condition, $t_\f-t_\i$ is equal to the interval $\Delta t_\mathrm{FS}=t_\f-t_\s$ between sample excitation by the signal photon and the subsequent fluorescence emission, enabling access to excited-state dynamics without any timing signal from the pump.
    }
    \label{fig2}
\end{figure}

Figure~\ref{fig1}(a) schematically illustrates the quantum spectroscopy setup proposed in the previous study. \cite{Fujihashi:2025tw}
As shown in Fig.~\ref{fig1}(a), the optical path length from the beam sampler to the nonlinear crystal was equated to the path length from the beam sampler to the photodiode. A pulsed laser pumped a type-II spontaneous parametric down-conversion (SPDC) crystal, generating entangled photon pairs. The beam sampler directed a fraction of the pump to the photodiode, triggering the time-correlated single-photon counting (TCSPC) timer. After leaving the crystal, a polarizing beam splitter separated the photon pairs. The idler photon was dispersed by a spectrometer and detected with a single-photon camera, such as a delay-line anode detector, \cite{Iso:2025ca,Iso:2025hy} thereby recording its frequency and arrival time. The signal photon excited a molecular sample under a microscope, and the resulting fluorescence was dispersed by a spectrometer and detected by a second single-photon camera. The TCSPC module registered the time correlations between the idler and the fluorescence photons. With the pump-pulse center defined as $t_\p=0$, and for entangled photon pairs exhibiting the negative temporal correlation $t_\s=-t_\i$, the elapsed time from excitation to fluorescence emission was calculated, $\Delta t_\mathrm{FS}=t_\f -t_\s=t_\f+t_\i$, as illustrated in Fig.~\ref{fig2}(a). Consequently, plotting the coincidence counts as a function of $t_\f+t_\i$ yielded a time-resolved two-dimensional fluorescence spectrum that tracked the excited-state dynamics of the sample.

Figure~\ref{fig1}(b) shows the configuration used in this study. Unlike the configuration shown in Fig.~\ref{fig1}(a), the arrival time of the laser was not recorded; instead, the arrival time of the idler photon, registered by the single-photon camera, defined the temporal reference, and the TCSPC module measured the delay between the two single-photon cameras $t_\f-t_\i$. This change simplified the optical system relative to that shown in Fig.~\ref{fig1}(a). As the timing of the pump pulse was not required, this configuration permitted the use of a monochromatic laser instead of a pulsed pump. Figure~\ref{fig2}(b) illustrates the corresponding detection scheme for negatively correlated frequency-photon pairs. As discussed in Sec.~IV, under the considered condition, the measured delay $t_\f - t_\i$ was equated to the interval between excitation of the sample by the signal photon and the subsequent fluorescence emission, $\Delta t_\mathrm{FS}=t_\f-t_\s$, enabling time-resolved observation of excited-state dynamics without any timing signal from the pump laser.

\section{Quantum states of entangled photons}
We demonstrated that time-resolved two-dimensional fluorescence spectra can be retrieved by referencing the detection time of the idler photons rather than the arrival time of the pump pulse. Contrary to the previous study, \cite{Fujihashi:2025tw} we formulate the quantum state of the light generated via the SPDC without fixing the central time of the pump pulse to zero; instead, it was treated as a free parameter $t_\p$.

\begin{figure}
    \centering
    \includegraphics{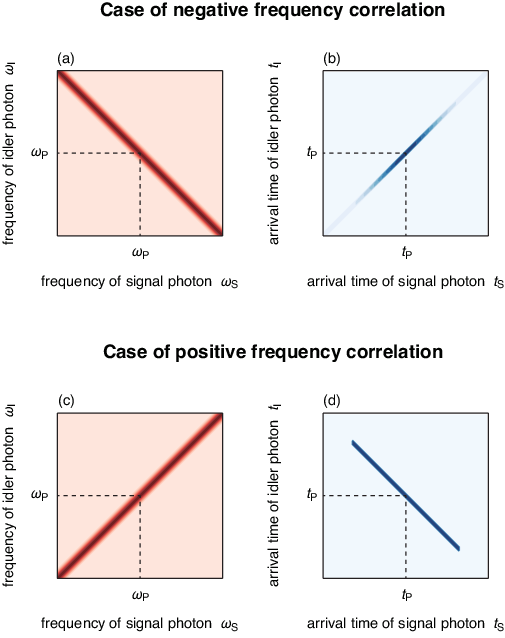}
    \caption{Joint spectral intensity $|f(\omega_\s,\omega_\i)|^2$ and the corresponding joint temporal intensity $|f(t_\s,t_\i)|^2$ of the two-photon state. Panels~(a) and (b) show the cases of negative frequency correlation, whereas panels~(c) and (d) show the cases of positive frequency correlation. The joint spectral intensities in panels~(a) and (c) are computed using Eqs.~\eqref{eq:psi-twin-negative} and \eqref{eq:psi-twin-positive}, respectively, while the joint temporal intensities in panels~(b) and (d) are based on Eqs.~\eqref{eq:temporal1} and \eqref{eq:temporal2}, respectively. For clarity, the delta functions in panels~(b) and (d) are rendered as finite-width rectangular functions.
    }
    \label{fig3}
\end{figure}

Let $\adagger_\s(\omega)$ and $\adagger_\i(\omega)$ denote the creation operators of the signal and idler photons of frequency $\omega$, respectively. The operators satisfy the commutation relation $[\a_\sigma(\omega),\adagger_{\sigma'}(\omega')] = \delta_{\sigma\sigma'}\,\delta(\omega-\omega')$.
In the weak down-conversion regime, the quantum state of light can be expressed as
\begin{align}
    \lvert \psi_\twin \rangle 
    = 
	\iint d\omega_1 d\omega_2 \,
    f(\omega_1,\omega_2)
    \adagger_\s(\omega_1) \adagger_\i(\omega_2)
    \lvert \vac \rangle,
    \label{eq:psi-twin}
\end{align}
where $f(\omega_1,\omega_2)$ denotes the two-photon amplitude expressed as 
\begin{align}
    f(\omega_1,\omega_2)
    &=
    \zeta\,
    e^{i(\omega_1+\omega_2)t_\p}
    \alpha_\p(\omega_1+\omega_2) 
\notag \\
    &\quad\times	
    \sinc\frac{\Delta k(\omega_1,\omega_2)L}{2},
    \label{eq:two-photon-amplitude}
\end{align}
where $\zeta$ corresponds to the conversion efficiency of the SPDC process, $\alpha_\p(\omega)$ denotes the normalized pump envelope function, $\Delta k(\omega_1,\omega_2)$ represents the wave vector mismatch between the input and output photons, $t_\p$ denotes the central time of the pump laser, and $L$ denotes the crystal length. 
Contrary to the previous study, \cite{Fujihashi:2025tw} an additional phase term $e^{i(\omega_1+\omega_2)t_\p}$ is included to Eq.~\eqref{eq:two-photon-amplitude}, accounting for the central time of the pump laser $t_\p$, as presented in the temporal distribution of the two-photon state in Figs.~\ref{fig3}(b) and \ref{fig3}(d). Here, Eq.~\eqref{eq:two-photon-amplitude} is derived from Appendix~A. In this study, the pump envelope function was modeled as a Gaussian function, defined by
\begin{align}
    \alpha_\p(\omega) 
    =
    \frac{1}{\sqrt{2\pi\sigma_\p^2}}
    \exp\left[-\frac{(\omega-\omega_\p)^2}{2\sigma_\p^2}\right],
\end{align}
where $\omega_\p$ denotes the central frequency of the pump laser. 
The wave vector mismatch in Eq.~\eqref{eq:two-photon-amplitude} can be approximated as \cite{Dorfman:2016da} $\Delta k(\omega_1,\omega_2)L=(\omega_1-\omega_\p/2)T_\s +(\omega_2-\omega_\p/2)T_\i$ with $T_\lambda=(v_\p^{-1}-v_\lambda^{-1})L$, where $v_\p^{-1}$ is the group velocity of the pump laser at the central frequency $\omega_\p$, and $v_\s^{-1}$ and $v_\i^{-1}$ denote the group velocities of the signal and idler photons at central frequency $\omega_\p/2$, respectively. This approximation remains valid for degenerate type-II down-conversion. The entanglement time $T_\e=|T_\s-T_\i|$ is used to quantify the spectral bandwidth of the two-photon amplitude.

In this study, we examined two classes of frequency correlations between photon pairs: positive and negative. As presented in Figs.~\ref{fig3}(a) and \ref{fig3}(c), a positive correlation corresponds to a joint spectrum elongated along $\omega_\s \approx \omega_\i$, whereas a negative correlation corresponds to the spectra constrained mainly by energy conservation, $\omega_\s+\omega_\i \approx \omega_\p$, producing an anti-diagonal elongation. Within the short entanglement time, $T_\e \ll \sigma_\p^{-1}$, Eq.~\eqref{eq:psi-twin} can be reduced to
\begin{align}
    \lvert \psi_\twin \rangle 
    &= 
    \zeta
	\iint d\omega_1 d\omega_2 \,
    e^{i(\omega_1+\omega_2)t_\p}
    \alpha_\p(\omega_1+\omega_2) 
\notag \\    
    &\quad\times    
    \adagger_\s(\omega_1) \adagger_\i(\omega_2)
    \lvert \vac \rangle.
    \label{eq:psi-twin-negative}
\end{align}
Thus, the two photons exhibit a negative frequency correlation. Based on the Fourier transformation relationship, the negative frequency correlation translates into a positive temporal correlation, as shown in Figs.~\ref{fig3}(a) and \ref{fig3}(b). As a representative of a positively frequency-correlated state, we consider the difference-beam state examined in the previous study, \cite{Fujihashi:2025tw} expressed as
\begin{align}
    \lvert \psi_\twin \rangle 
    &= 
    \zeta
	\iint d\omega_1 d\omega_2 \,
    e^{i(\omega_1+\omega_2)t_\p}    
    \phi(\omega_1-\omega_2) 
\notag \\    
    &\quad\times    
    \adagger_\s(\omega_1) \adagger_\i(\omega_2)
    \lvert \vac \rangle,
    \label{eq:psi-twin-positive}
\end{align}
where $\phi(\omega)=\sinc(\omega T_\e/4)$ denotes the phase-matching function. Here, Eq.~\eqref{eq:psi-twin-positive} can be obtained under degenerate type-II symmetric group-velocity matching, \cite{Ansari:2018ta} assuming impulsive pumping. Further, the positive temporal correlation or the negative frequency correlation was essential for various measurements in this study, considering the detection time of the idler photon as the temporal reference.

\section{Two-photon coincidence signal}

We investigated the time and frequency-resolved two-photon coincidence signals obtained from the measurements, as shown in Fig.~\ref{fig1}(b). The total Hamiltonian can be expressed as: $\hat{H} = \hat{H}_{\rm mol} + \hat{H}_{\rm field} + \hat{H}_{\rm mol-field}$, \cite{Fujihashi:2025tw} where the first term represents the Hamiltonian describing the photoactive degrees of freedom in the molecules, while the second term is the free Hamiltonian of the fields. Under the rotating-wave approximation, the molecule-field interaction can be written as: $\hat{H}_{\rm mol-field}(t)=-\hat{\mu}_+ \E_\s^+(t) -\hat{\mu}_+ \E_\mathrm{F}^+(t) + \mathrm{h.c.}$, where $\hat{\mu}_+$ denotes the transition dipole operator describing the optical transition from the electronic ground state to the electronic single excited state. The positive frequency component of the electric field operators before reaching the single-photon cameras can be expressed as
\begin{align}
    \E_\sigma^+(t)
    =
    \frac{1}{2\pi}
    \int d\omega
    \,
    \hat{a}_\sigma(\omega)e^{-i\omega t}.
    \label{eq:field}
\end{align}
The negative frequency component can be expressed as $\E_\sigma^-(t)=\E_\sigma^+(t)^\dagger$.

\begin{figure}
    \centering
    \includegraphics{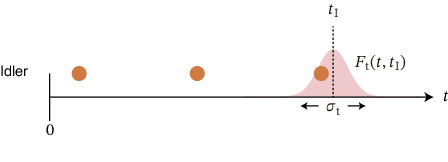}
    \caption{Schematic of photon detection events. The light-pink shaded region illustrates the temporal distribution of $F_\mathrm{t}(t,t_\i)$ in Eq.~\eqref{eq:time-gate-function}. The arrival times of the idler photons are blurred by the finite temporal resolution of the single-photon camera. The fluorescence photons experience the same temporal blurring.
    }
    \label{fig4}
\end{figure}

\begin{figure}
    \centering
    \includegraphics{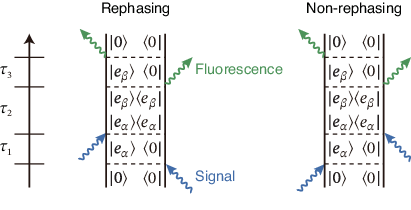}
    \caption{Double-sided Feynman diagrams representing the Liouville space pathways contributing to the two-photon coincidence signal in Eq.~\eqref{eq:tpc-signal2}.
    }
    \label{fig5}
\end{figure}

As described in Sec.~II and shown in Fig.~\ref{fig1}, we simultaneously resolved the frequency and arrival time of photons using a single-photon camera. Physically, the time-frequency-resolved photon-counting signal produced by the single-photon camera and spectrometer can be expressed as a two-time integral of the first-order temporal field correlation function, weighted by the instrument's temporal window and by the autocorrelation of the spectrometer response. \cite{Eberly:1977ti} To systematically extend this description to two-photon coincidence signals, the bare-field operators in Eq.~\eqref{eq:field} are substituted with operators that explicitly incorporate the temporal and spectral resolutions of the detector, \cite{Dorfman:2012no,Dorfman:2016da,Fujihashi:2025tw} such as the field operator measured by the single-photon camera. Accordingly, two functions are defined as \cite{Dorfman:2016da,Gelin:2002ti,Yang:2023two}
\begin{align}
	F_\mathrm{t}(t,t_a)
	=
	\exp\left[-\frac{1}{2\sigma_\mathrm{t}^2} (t -t_a)^2\right],
    \label{eq:time-gate-function}	
\end{align}
\begin{align}
	F_\mathrm{f}(\omega,\omega_a)
	=
    \frac{\sigma_\mathrm{f}}{i(\omega_a - \omega) + \sigma_\mathrm{f}}
	\label{eq:frequency-filter-function}	
\end{align}
for $a=\i$ and $\f$. Here, $\omega_\i$ and $\omega_\f$ denote the center frequencies of the idler and fluorescence photons, respectively, recorded by the single-photon camera, and $t_\i$ and $t_\f$ represent their corresponding arrival times recorded by the same camera. As shown in Fig.~\ref{fig4}, Eq.~\eqref{eq:time-gate-function} describes the uncertainty in photon arrival time associated with the temporal resolution of the single-photon camera. Similarly, Eq.~\eqref{eq:frequency-filter-function} describes the uncertainty in the detected frequency of photon~$a$ associated with the frequency resolution of the single-photon camera. The electric-field information obtained from the single-photon cameras can be expressed as \cite{Dorfman:2012no,Dorfman:2016da}
\begin{align}
	\hat{\mathcal{E}}_a^+(\omega_a,t_a;t)
	=	
    \int_{-\infty}^\infty ds
    \,	
    F_{\rm f}(t-s,\omega_a)
    F_\mathrm{t}(s,t_a)
	\hat{E}_a^+(s),
	\label{eq:gated-electric-field}
\end{align}
where $F_\mathrm{f}(t,\omega_a)$ is the Fourier transform of Eq.~\eqref{eq:frequency-filter-function}, $F_\mathrm{f}(t,\omega_a)=\theta(t) \sigma_\mathrm{f}\,  e^{-i \omega_a t - \sigma_\mathrm{f}t}$ with $\theta(t)$
being the Heaviside step function. Using Eq.~\eqref{eq:gated-electric-field}, the time and frequency-resolved two-photon coincidence signals can be derived as \cite{Dorfman:2012no,Fujihashi:2025tw}
\begin{widetext}
\begin{align}
    S(\omega_\f,t_\f;\omega_{\i},t_{\i})
	=
	\int_{-\infty}^\infty dt
	\int_{-\infty}^\infty ds
	\,
	\tr\left[  
	   \hat{\mathcal{E}}_\f^{-}(\omega_\f,t_\f;t)
	   \hat{\mathcal{E}}_\f^{+}(\omega_{\rm F},t_\f;t)
	   \hat{\mathcal{E}}_{\i}^{-}(\omega_{\i},t_{\i};s)
	   \hat{\mathcal{E}}_{\i}^{+}(\omega_{\i},t_{\i};s)
	   \hat{\rho}(t)     
    \right]   
	\label{eq:tpc-signal} 
\end{align}
\end{widetext}
with the initial condition of $\hat{\rho}(-\infty) = \hat{\rho}_\mathrm{mol}^\mathrm{eq} \otimes \lvert \psi_\twin \rangle\langle \psi_\twin \rvert$, where $\hat{\rho}_\mathrm{mol}^\mathrm{eq}$ represents the thermal equilibrium state of the photoactive degrees of freedom in the molecules. A perturbative expansion of the density operator $\hat{\rho}(t)$ in the molecule-field interaction revealed that the lowest non zero contribution arose at third order. As shown in Fig.~\ref{fig5}, the resulting signal comprised two components: rephasing and non-repasing stimulated emission (SE) pathways. \cite{Mukamel:1995us} Accordingly, Eq.~\eqref{eq:tpc-signal} can be written as the sum of the rephasing and non-rephasing contributions, given by
\begin{align}
    S(\omega_\f,t_\f;\omega_{\i},t_{\i})
	=
    \sum_{x=\mathrm{r,nr} }
    S^{(x)}(\omega_\f,t_\f;\omega_{\i},t_{\i}),
	\label{eq:tpc-signal2} 
\end{align}
in which each contribution can be expressed as
\begin{widetext}
\begin{align}
    S^{(x)}(\omega_\f,t_\f;\omega_{\i},t_{\i})
    &=
	\int_{-\infty}^\infty dt
	\int_{0}^\infty d\tau_3    
	\int_{0}^\infty d\tau_2    
	\int_{0}^\infty d\tau_1 
	\int_{-\infty}^\infty ds
	\int_{-\infty}^s ds'
	\,
    F_\mathrm{t}(t,t_\f)
    F_\mathrm{t}(t-\tau_3,t_\f)
    F_\mathrm{t}(s,t_\i)
    F_\mathrm{t}(s',t_\i)
\notag \\
    &\quad\times
    e^{i(\omega_\f+i\sigma_\mathrm{f})\tau_3}
    \Phi_\SE^{(x)}(\tau_3,\tau_2,\tau_1)
    C^{(x)}(\omega_\f, \omega_\i,t,s,s';\tau_3,\tau_2,\tau_1).
	\label{eq:tpc-signal3} 
\end{align}
The function $\Phi_\SE^{(x)}(\tau_3,\tau_2,\tau_1)$ denotes the third-order response function of molecules, and $C^{(x)}(\omega_\f, \omega_\i, t, s, s';\allowbreak \tau_3,\tau_2,\tau_1)$ represents the four-body correlation function of the electric field operators as
\begin{align}
    C^{\rephasing}(\omega_\f, \omega_\i,t,s,s';\tau_3,\tau_2,\tau_1)
    &=
    \langle \psi_\twin \rvert
        [
        e^{i(\omega_\i+i\sigma_\mathrm{f})(s-s')}    
        \E_\i^-(s')\E_\i^+(s) + \mathrm{h.c.}
        ]
\notag \\
    &\quad\times
        \E_\s^-(t-\tau_3-\tau_2-\tau_1)
        \E_\s^+(t-\tau_3-\tau_2)
    \lvert \psi_\twin \rangle,
\end{align}
\begin{align}
    C^{\nonrephasing}(\omega_\f, \omega_\i,t,s,s';\tau_3,\tau_2,\tau_1)
    &=
    \langle \psi_\twin \rvert
        [
        e^{i(\omega_\i+i\sigma_\mathrm{f})(s-s')}    
        \E_\i^-(s')\E_\i^+(s) + \mathrm{h.c.}
        ]
\notag \\
    &\quad\times
        \E_\s^-(t-\tau_3-\tau_2)
        \E_\s^+(t-\tau_3-\tau_2-\tau_1)
    \lvert \psi_\twin \rangle.
\end{align}
\end{widetext}
To obtain explicit expressions for the signal, two limiting cases were analyzed, where the frequencies of the entangled pairs were either negatively or positively correlated.

\subsection{Negative frequency correlations}

We considered the quantum state of light, as expressed by Eq.~\eqref{eq:psi-twin-negative}. In this case, the two-photon wavefunction can be expressed as
\begin{align}
    \langle \vac \rvert
    \E_\s^+(t)
    \E_\i^+(s)
    \lvert \psi_\twin \rangle 
    =
    \zeta \,
    \delta(t-s)
    A_\p(t-t_\p)
    e^{-i\omega_\p(t-t_\p)}.
	\label{eq:temporal1} 
\end{align}
In Eq.~\eqref{eq:temporal1}, all terms except the delta function represent the pump envelope in the time domain, obtained from the Fourier transform of the spectral profile $\alpha_\p(\omega)$. The phase-free magnitude of this envelope can be determined as
\begin{align}
    A_\p(t)
    =
    \frac{1}{2\pi}
    \exp
    \left(-\frac{\sigma_\p^2 t^2}{2}\right).
\end{align}
Consequently, the rephasing signal in Eq.~\eqref{eq:tpc-signal3} can be written as 
\begin{widetext}
\begin{multline}
    S^{\rephasing}(\omega_\f,t_\f;\omega_{\i},t_{\i})
    =
    \frac{\zeta^2}{2}    
    \Re\,
	\int_{-\infty}^\infty dt
	\int_{0}^\infty d\tau_3    
	\int_{0}^\infty d\tau_2    
	\int_{0}^\infty d\tau_1 
    e^{i(\omega_\f+i\sigma_\mathrm{f})\tau_3}
    e^{-i(\omega_\p-\omega_\i-i\sigma_\mathrm{f})\tau_1}    
\\    
    \times     
    F_\mathrm{t}(t,t_\f)
    F_\mathrm{t}(t-\tau_3,t_\f)
    F_\mathrm{t}(t-\tau_3-\tau_2,t_\i) 
    F_\mathrm{t}(t-\tau_3-\tau_2-\tau_1,t_\i)   
\\    
    \times      
    A_\p(t-\tau_3-\tau_2-t_\p)
    A_\p(t-\tau_3-\tau_2-\tau_1-t_\p)
    \Phi_\SE^{\rephasing}(\tau_3,\tau_2,\tau_1). 
	\label{eq:tpc-signal-negative0}     
\end{multline}
\end{widetext}
Assuming that the time resolution of the detector is sufficiently short compared with the timescales over which the molecular response and the pump envelope vary, we approximate one of the two functions $F_\mathrm{t}(t,t_a)$ associated with fluorescence detection and one of the two associated with idler-photon detection by delta functions, $F_\mathrm{t}(t,t_a)\to \delta(t-t_a)$. Equation~\eqref{eq:tpc-signal-negative0} then simplifies to
\begin{align}
    S^{\rephasing}&(\omega_\f,t_\f;\omega_{\i},t_{\i})
    =
\notag \\    
    &
    \frac{\zeta^2}{2}
    \Re\,
    \hat{\mathrm{F}}_1
    \left[
    \mathcal{A}(\tau_1)
    \mathcal{F}(\tau_3,\tau_1)
    \Phi_\SE^{\rephasing}(\tau_3,t_\f-t_\i-\tau_1,\tau_1)    
    \right],
	\label{eq:tpc-signal-negative1} 
\end{align}
where $\mathcal{F}(\tau_3,\tau_1)$ and $\mathcal{A}(\tau_1)$ are defined as
\begin{align}
    \mathcal{F}(\tau_3,\tau_1)
    =
    F_\mathrm{t}(t_\f+\tau_3,t_\f)
    F_\mathrm{t}(t_\i+\tau_1,t_\i),
\end{align}
\begin{align}
    \mathcal{A}(\tau_1)
    =
    A_\p(\tau_1+t_\i-t_\p)
    A_\p(t_\i-t_\p).
\end{align}
The operator $\hat{\mathrm{F}}_1$ can be defined for any function $f(\tau_3,\tau_1)$ as
\begin{align}
    \hat{\mathrm{F}}_1 [f(\tau_3,\tau_1)]
    &=
    \int_0^\infty d\tau_3
    \,
    e^{i(\omega_\f+i\sigma_\mathrm{f})\tau_3}
\notag \\
    &\quad\times 
    \int_0^\infty d\tau_1
    \,
    e^{\mp i(\omega_\p-\omega_\i \mp i\sigma_\mathrm{f})\tau_1}
    f(\tau_3,\tau_1).
	\label{eq:operator-f1} 
\end{align}
In Eq.~\eqref{eq:operator-f1}, the sign $\mp$ can be determined by the Liouville pathway: the upper sign ($-$) corresponds to the rephasing contribution, whereas the lower sign ($+$) corresponds to the non-rephasing contribution. Considering the limit of monochromatic pumping with frequency $\omega_\p$, in which the pump envelope function reduces to $A_\p(t)=1$ and $\mathcal{A}(t)=1$, Eq.~\eqref{eq:tpc-signal-negative1} can be rewritten as
\begin{align}
    S^{\rephasing}&(\omega_\f,t_\f;\omega_{\i},t_{\i})
    =
\notag \\    
    &
    \frac{\zeta^2}{2}
    \Re\,
    \hat{\mathrm{F}}_1
    \left[
    \mathcal{F}(\tau_3,\tau_1)
    \Phi_\SE^{\rephasing}(\tau_3,t_\f-t_\i-\tau_1,\tau_1)    
    \right]. 
	\label{eq:tpc-signal-negative2} 
\end{align}
Similarly, the non-rephasing signal can be calculated. In this study, we assumed that the dynamics within the single-excitation manifold were sufficiently slow compared to the decay of coherence between different manifolds during the $\tau_1$ period. Therefore, $t_\f-t_\i-\tau_1$ in Eq.~\eqref{eq:tpc-signal-negative2} can be approximated as $t_\f-t_\i$, and consequently, Eq.~\eqref{eq:tpc-signal-negative2} can be simplified to
\begin{align}
    S^{(x)}(\omega_\f,t_\f;\omega_{\i},t_{\i})
    \simeq
    \frac{\zeta^2}{2}
    \Re\,
    \hat{\mathrm{F}}_1
    \left[
    \mathcal{F}(\tau_3,\tau_1)
    \Phi_\SE^{(x)}(\tau_3,t_\f-t_\i,\tau_1)    
    \right]. 
	\label{eq:tpc-signal-negative3} 
\end{align}
Hence, excited-state dynamics in a molecular system can be monitored in real time using idler-photon detection time as the temporal reference and recording $t_\f-t_\i$, without measuring the pump-laser arrival time. 
Concurrently, the proposed method provides spectral information equivalent to the SE contribution of the two-dimensional (2D) photon echo spectrum, \cite{Brixner:2005wu,Fujihashi:2015kz} expressed as a function of frequencies $\omega_\p-\omega_\i$ and $\omega_\f$, as indicated in Eqs.~\eqref{eq:operator-f1} and \eqref{eq:tpc-signal-negative3}, respectively.

\subsection{Positive frequency correlations}

Considering the quantum state of light, as in Eq.~\eqref{eq:psi-twin-positive}, the two-photon wave function can be calculated as
\begin{align}
    \langle \vac \rvert
    \E_\s^+(t)
    \E_\i^+(s)
    \lvert \psi_\twin \rangle 
    =
    \zeta \,
    D(s-t_\p)
    \delta(t+s-2t_\p),
	\label{eq:temporal2} 
\end{align}
where $D(t)$ is the Fourier transform of the phase matching function,
\begin{align}
    D(t)
    =
    \frac{1}{2\pi}
    \int_{-\infty}^\infty d\omega 
    \,
    \phi(\omega)
    e^{-i\omega t}.
	\label{eq:PMF-Fourier} 
\end{align}
Thus, the rephasing signal in Eq.~\eqref{eq:tpc-signal3} can be expressed as
\begin{align}
    S^{\rephasing}&(\omega_\f,t_\f;\omega_{\i},t_{\i})
    =
\notag \\    
    &
    \frac{\zeta^2}{2}
    \Re\,
    \hat{\mathrm{F}}_2
    \left[
    \mathcal{D}(\tau_1)
    \mathcal{F}(\tau_3,\tau_1)
    \Phi_\SE^{\rephasing}(\tau_3,t_\f+t_\i-2t_\p,\tau_1)    
    \right]    
	\label{eq:tpc-signal-positive1} 
\end{align}
with
\begin{align}
    \mathcal{D}(\tau)
    =
    D(t_\p-t_\i-\tau)
    D(t_\p-t_\i).
\end{align}
As in the previous subsection, the derivation of Eq.~\eqref{eq:tpc-signal-positive1} assumed that the time resolution of the detector is sufficiently shorter than both the timescale of the system dynamics and the entanglement time.
Unlike the transform used in Sec.~IVA, the Fourier-Laplace transform in Eq.~\eqref{eq:tpc-signal-positive1} is independent of the central frequency of the pulse laser $\omega_\p$.
Accordingly, the operator $\hat{\mathrm{F}}_1$ defined in Eq.~\eqref{eq:operator-f1} is replaced by a new operator $\hat{\mathrm{F}}_2$ for any function $f(\tau_3,\tau_1)$ as
\begin{align}
    \hat{\mathrm{F}}_2 [f(\tau_3,\tau_1)]
    &=
    \int_0^\infty d\tau_3
    \,
    e^{i(\omega_\f+i\sigma_\mathrm{f})\tau_3}
\notag \\
    &\quad\times 
    \int_0^\infty d\tau_1
    \,
    e^{\mp i(\omega_\i \mp i\sigma_\mathrm{f})\tau_1}
    f(\tau_3,\tau_1).
	\label{eq:operator-f2} 
\end{align}
As in Eq.~\eqref{eq:operator-f1}, $\mp$ in Eq.~\eqref{eq:operator-f2} can be determined using the Liouville pathway. To consider the case of a strong positive frequency correlation, the limit of a long entanglement time was considered. Within this limit, Eq.~\eqref{eq:PMF-Fourier} yields $D(t)=1$ and thus $\mathcal{D}(t)=1$. Consequently, Eq.~\eqref{eq:tpc-signal-positive1} can be rewritten as
\begin{align}
    S^{(x)}&(\omega_\f,t_\f;\omega_{\i},t_{\i})
    =
\notag \\    
    &
    \frac{\zeta^2}{2}
    \Re\,
    \hat{\mathrm{F}}_2
    \left[
    \mathcal{F}(\tau_3,\tau_1)
    \Phi_\SE^{(x)}(\tau_3,t_\f+t_\i-2t_\p,\tau_1)    
    \right].    
	\label{eq:tpc-signal-positive2} 
\end{align}
Unlike that in Eq.~\eqref{eq:tpc-signal-negative3}, the response function in Eq.~\eqref{eq:tpc-signal-positive2} depends on the pump-laser center time $t_\p$, not on the fluorescence-idler detection time difference, $t_\f-t_\i$. Consequently, the measurement scheme presented in Fig.~\ref{fig1}(b) cannot be implemented in the case of positive frequency correlations.

\subsection{Uncorrelated states}

In the previous subsection, we examined photon pairs that exhibited positive and negative frequency correlations. However, even when the biphoton state itself is spectrally correlated, on time scales relevant to the measurement, the frequency and temporal degrees of freedom can effectively behave as if they were uncorrelated. Therefore, we analyzed the case and, as a representative, considered time ordering
\begin{align}
    \sigma_\p^{-1}, \, T_\e 
    \ll
    \sigma_\mathrm{t}
    \ll
    \tau_\mathrm{ex},
\end{align}
where $\tau_\mathrm{ex}$ denotes the characteristic timescale of molecular dynamical processes. Accordingly, (i) the temporal width of the two-photon distribution was negligibly small compared with other characteristic times, and (ii) the difference between $\sigma_\p^{-1}$ and $T_\e$ was experimentally insignificant. Thus, the temporal distribution was regarded as being effectively uncorrelated. Consequently, the two-photon wavefunction can be approximated as the product of delta functions as
\begin{align}
    \langle \vac \rvert
    \E_\s^+(t)
    \E_\i^+(s)
    \lvert \psi_\twin \rangle 
    \simeq
    \zeta \,
    \delta(s-t_\p)
    \delta(t-t_\p).
	\label{eq:temporal3} 
\end{align}
By substituting Eq.~\eqref{eq:temporal3} into Eq.~\eqref{eq:tpc-signal3} and repeating the relevant calculations, the signal defined in Eq.~\eqref{eq:tpc-signal3} can be reduced to
\begin{align}
    S^{(x)}&(\omega_\f,t_\f;\omega_{\i},t_{\i})
    \simeq
    \delta(t_\p-t_\i)
    \frac{\zeta^2}{2}
    \Re\,
    \int_0^\infty d\tau_3
    \,
    e^{i(\omega_\f+i\sigma_\mathrm{f})\tau_3}
\notag \\
    &\quad\times    
    \mathcal{F}(\tau_3,0)
    \Phi_\SE^{(x)}(\tau_3,t_\f-t_\i,0).
	\label{eq:tpc-signal-uncorrelated} 
\end{align}
The detection scheme in Fig.~\ref{fig1}(b) yielded a time-resolved fluorescence spectrum even when the entangled photons were effectively uncorrelated in frequency, as expressed by Eq.~\eqref{eq:tpc-signal-uncorrelated}. However, within this limit, the idler frequency recorded by the single-photon camera carried no information about the frequency of the signal photon; therefore, the spectroscopic result reduced to a one-dimensional spectrum along the fluorescence-photon frequency axis. This behavior was consistent with the calculated results of the previous study. \cite{Fujihashi:2025tw} Under these circumstances, replacing the pump laser with entangled photons offered no simplification over conventional time-resolved fluorescence apparatus.

\begin{figure*}
    \centering
    \includegraphics{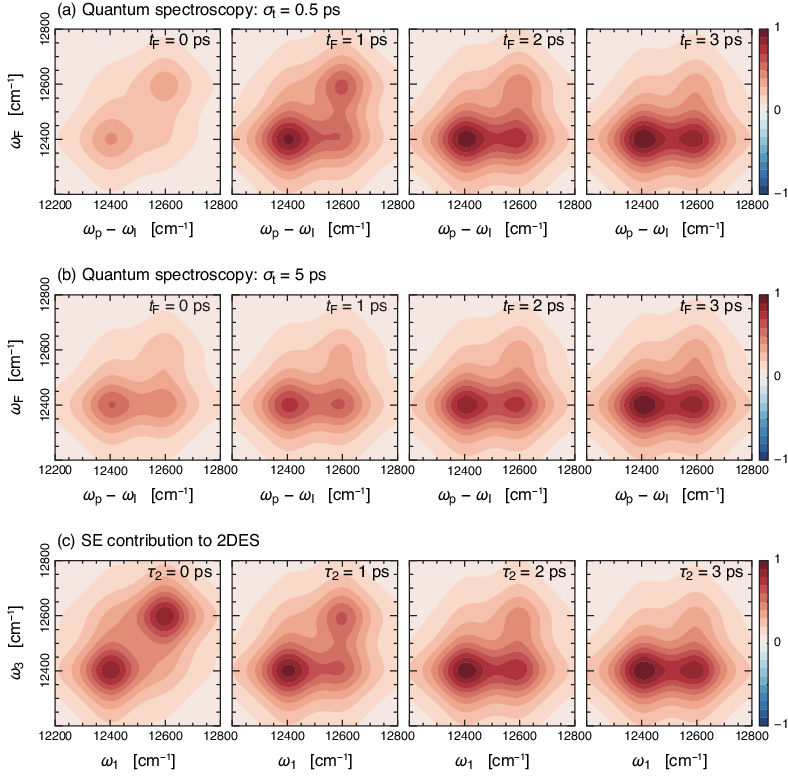}
    \caption{Time evolution of the 2D spectra obtained from the quantum spectroscopic scheme described by Eq.~\eqref{eq:tpc-signal-negative0} for (a) $\sigma_\mathrm{t}=0.5\,\mathrm{ps}$ and (b) $\sigma_\mathrm{t}=5\,\mathrm{ps}$. The pump bandwidth and central frequency are set to $\sigma_\p=10\,\mathrm{cm}^{-1}$ and $\omega_\p=25000\,\mathrm{cm^{-1}}$, respectively, where $\sigma_\p^{-1}\approx 3.34\,\mathrm{ps}$. 
    The frequency resolution of the spectrometer is set to $\sigma_\mathrm{f}=0$, and the detection time of the idler photon is fixed at $t_\i=0$. (c) Time evolution of the stimulated-emission (SE) contribution to the 2D electronic spectrum. The contour plots in panels (a)--(c) are normalized such that the maximum value of each spectrum at $t_\mathrm{F}=3\,\mathrm{ps}$ ($\tau_2=3\,\mathrm{ps}$) is unity, and equally spaced contour levels ($0, \pm 0.1, \pm 0.2, \dots$) are shown.
    }    
    \label{fig6}
\end{figure*}

\section{Numerical results}

In deriving Eq.~\eqref{eq:tpc-signal-negative1} from Eq.~\eqref{eq:tpc-signal-negative0}, we assumed that the detector time resolution $\sigma_\mathrm{t}$ is much shorter than both the characteristic timescale $\tau_\mathrm{ex}$ of the molecular dynamics and the pump-envelope timescale $\sigma_\p^{-1}$. Under this assumption, two of the four functions $F_\mathrm{t}(t,t_a)$ appearing in Eq.~\eqref{eq:tpc-signal-negative0} were approximated by delta functions. The same approximation was also used in the derivation of Eq.~\eqref{eq:tpc-signal-positive1} for positively frequency-correlated entangled photon pairs. In Ref.~\onlinecite{Fujihashi:2025tw}, the 2D spectra were calculated using these approximate expressions only in the parameter regime satisfying $\sigma_t \ll \tau_{\mathrm{ex}}, T_\e$. This approximation is mathematically rather crude, because in the formal limit of sufficiently short detector time resolution all four $F_t(t,t_a)$ terms approach delta functions, whereas only two of them are replaced by delta functions while the other two are kept finite. It is therefore important to assess the validity of this approximation by performing numerical calculations based on the more general expression.

In this section, we perform such calculations using the full expression in Eq.~\eqref{eq:tpc-signal-negative0}. Specifically, we verify that the statement of Eq.~\eqref{eq:tpc-signal-negative3}, namely that the excited-state dynamics can be monitored through the time difference $t_F-t_I$, is valid in the regime $\sigma_t \ll \tau_{\mathrm{ex}}, \sigma_p^{-1}$, and we examine how finite detector time resolution affects the resulting spectroscopic signals.
In what follows, we consider a coupled dimer as a model system for the electronic excitations. For the numerical calculations, we employ a simplified form of the molecular response function. Specifically, we neglect memory effects that extend across different time intervals in the response function,\cite{Ishizaki:2020jl} and also ignore coherences between the electronic eigenstates in the single-excitation manifold. Under these assumptions, the response function is written in the factorized form
\begin{align}
    \Phi_{\SE}^{(x)}(\tau_3,\tau_2,\tau_1)
    =
    \sum_{\alpha,\beta=1}^{2}
    \mu_{\beta 0}^{2}\mu_{\alpha 0}^{2}
    G_{\beta 0}(\tau_3)
    G_{\beta\beta\leftarrow\alpha\alpha}(\tau_2)
    G_{x,\alpha}(\tau_1),
\end{align}
where $\mu_{\alpha 0}$ denotes the transition dipole moment of exciton state $\alpha$, and
\begin{align}
    G_{x,\alpha}(\tau_1)
    =
    \begin{cases}
        G_{0\alpha}(\tau_1), & x=\rephasing,\\[4pt]
        G_{\alpha 0}(\tau_1), & x=\nonrephasing.
    \end{cases}
\end{align}
The time evolution in the $\tau_1$ and $\tau_3$ periods is modeled phenomenologically as $G_{\alpha\beta}(t)=e^{-(i\omega_{\alpha\beta}+\Gamma_{\mathrm{env}})t}$, and the population transfer during the waiting time is taken to be \cite{Schlawin:2016er}
\begin{align}
    G_{\beta\beta\leftarrow\alpha\alpha}(t)
    =
    \theta(t)
    \begin{pmatrix}
        e^{-t/\tau_{\mathrm{ex}}} & 0 \\
        1-e^{-t/\tau_{\mathrm{ex}}} & 1
    \end{pmatrix}.
\end{align}
Here, the population of exciton state 1 decays with the time constant $\tau_{\mathrm{ex}}$, while that of exciton state 2 is fed by the transfer from state 1.
For numerical calculations, we chose the following parameters of the molecular system: $\omega_{10}=12600\,\mathrm{cm^{-1}}$, $\omega_{20}=12400\,\mathrm{cm^{-1}}$, $\Gamma_\mathrm{env}=100\,\mathrm{cm}^{-1}$, and $\tau_\mathrm{ex}=2\,\mathrm{ps}$.
For comparison, we also consider the SE contribution to the 2D electronic spectrum, whose definition is given in Appendix~\ref{appendixB}.

Figures~\ref{fig6}(a) and (b) present the 2D spectra obtained from the quantum spectroscopic scheme described by Eq.~\eqref{eq:tpc-signal-negative0} for $\sigma_\mathrm{t}=0.5\,\mathrm{ps}$ and $5\,\mathrm{ps}$, respectively.  
We set the pump bandwidth and central frequency to $\sigma_\p=10\,\mathrm{cm}^{-1}$ and $\omega_\p=25000\,\mathrm{cm^{-1}}$, respectively, where $\sigma_\p^{-1}\approx 3.34\,\mathrm{ps}$. We also set $\sigma_\mathrm{f}=0$ and fix the idler-photon detection time at $t_\i=0$.
Since the photon pairs in Eq.~\eqref{eq:psi-twin-negative} exhibit a negative frequency correlation satisfying $\omega_\s+\omega_\i\approx\omega_\p$, we plot the horizontal axis as $\omega_\p-\omega_\i$ rather than $\omega_\i$ itself, to facilitate direct comparison with the SE contribution to the 2D electronic spectrum presented in Fig.~\ref{fig6}(c).

As shown in Fig.~\ref{fig6}(a), two diagonal peaks appear in the 2D spectrum at $t_\f=0$. As time evolves, population transfer from state 1 to state 2 causes the diagonal peak of exciton state 1 to decay, while a cross peak emerges. At $t_\f=1$, $2$, and $3\,\mathrm{ps}$, the spectral profiles in Fig.~\ref{fig6}(a) closely agree with those of the 2D electronic spectra shown in Fig.~\ref{fig6}(c). This demonstrates that, in a time range sufficiently longer than $\sigma_\mathrm{t}$, the excited-state dynamics can indeed be monitored in real time, as predicted by Eq.~\eqref{eq:tpc-signal-negative3}. At $t_\f=0$, although the overall spectral shape is similar, the signal intensity is smaller than that of the 2D electronic spectrum.
This difference arises because the calculation of the 2D electronic spectrum assumes impulsive pulses, whereas in the quantum spectroscopic scheme the finite detector time resolution blurs the arrival time of the idler photon. As a result, the excitation time of the corresponding signal photon interacting with the molecular system can only be inferred with limited temporal precision.
As shown in Fig.~6(b), this effect becomes much more pronounced when $\sigma_t$ exceeds the characteristic timescale $\tau_{\mathrm{ex}}$ of the excited-state dynamics. In this case, the diagonal peak of state 1 has already disappeared and a cross peak has emerged at $t_F=0$, indicating that the excited-state dynamics can no longer be tracked accurately from the spectroscopic signal. These results confirm the validity of Eq.~\eqref{eq:tpc-signal-negative3} in the regime $\sigma_t \ll \tau_{\mathrm{ex}}, \sigma_p^{-1}$, while also showing that finite detector time resolution limits the ability of the present scheme to resolve excited-state dynamics accurately.

\begin{table*}
  \centering
  \renewcommand{\arraystretch}{1.4}
  \caption{Summarizing the influence of frequency correlation between twin photons on the feasibility of the measurement scheme, as shown in Fig.~\ref{fig1}(b).}
  \begin{tabular*}{\textwidth}{@{\extracolsep{\fill}}lccc}
    \hline\hline
    & \shortstack[c]{Negative frequency\\correlations}
    & \shortstack[c]{Positive frequency\\correlations}
    & \shortstack[c]{Uncorrelated\\states} \\    
    \hline
    Photon state
      & $T_\e \ll \sigma_\p^{-1}$
      & $\sigma_\p^{-1} \ll T_\e$
      & $\sigma_\p^{-1} \ll T_\e$ or $T_\e \ll \sigma_\p^{-1}$ \\
    Timescale of measurements
      & \shortstack[c]{$T_\e \ll \sigma_\mathrm{t} \ll \tau_\mathrm{ex},\,\sigma_\p^{-1}$}
      & \shortstack[c]{$\sigma_\p^{-1} \ll \sigma_\mathrm{t} \ll \tau_\mathrm{ex},\,T_\e$}
      & \shortstack[c]{$\sigma_\p^{-1},\, T_\e \ll \sigma_\mathrm{t} \ll \tau_\mathrm{ex}$} \\
    Limiting cases 
      & \shortstack[c]{CW pumping:\\$T_\e \ll \sigma_\mathrm{t} \ll \tau_\mathrm{ex} \ll \sigma_\p^{-1}$}
      & \shortstack[c]{Long entanglement time:\\$\sigma_\p^{-1} \ll \sigma_\mathrm{t} \ll \tau_\mathrm{ex} \ll T_\e$}
      & --- \\
    Feasibility of scheme in Fig.~\ref{fig1}(b)
      & Possible
      & Impossible
      & Possible \\
    Spectroscopic dimensionality
      & Two-dimensional
      & Two-dimensional
      & One-dimensional only \\
    Theory
      & Eq.~\eqref{eq:tpc-signal-negative3}
      & Eq.~\eqref{eq:tpc-signal-positive2}, Ref.~\onlinecite{Fujihashi:2025tw}
      & Eq.~\eqref{eq:tpc-signal-uncorrelated} \\
    \hline\hline
  \end{tabular*}
  \label{table1}
\end{table*}

\section{Conclusions}

In this study, we theoretically investigated the feasibility of applying an idler-referenced timing scheme that uses the arrival of an idler photon as a temporal reference in time-resolved two-dimensional fluorescence spectroscopy. The main findings of this study are summarized in Table~I. We demonstrated that the proposed scheme is viable when the entangled photon pair exhibits either negative or negligible frequency correlations. This approach simplifies the optical layout and lowers the experimental barrier to implementation by eliminating the pulse-timing detection stage required in previous setups. Furthermore, our analysis revealed that although entangled photons may exhibit frequency correlations in isolation, their frequency-time degrees of freedom can behave as effectively uncorrelated when considered over the full measurement timescale. To fully harness non-classical frequency-time correlations in time-resolved two-dimensional fluorescence spectroscopy, the temporal characteristics of the entangled photon source must therefore be carefully designed and matched to the overall timescale of the experiment.

\begin{acknowledgments}
A.I. is grateful to Dr.~Shuntaro~Tani and Prof.~Koichiro~Tanaka for their inspiring discussions. 
This study was supported by the MEXT Quantum Leap Flagship Program (Grant Number~JPMXS0118069242) and JSPS KAKENHI (Grant Number~JP21H01052). Y.F. acknowledges the support from JSPS KAKENHI (Grant Number~JP23K03341).
\end{acknowledgments}

\section*{Author declarations}

\subsection*{Conflict of interest}
The authors have no conflicts to disclose.

\subsection*{Author contributions}
{\bf Yuta Fujihashi}:
Conceptualization (equal);
Investigation (lead);
Writing - original draft (equal); 
Writing - review \& editing (equal).

{\bf Akihito Ishizaki}:
Conceptualization (equal); 
Funding acquisition (lead); 
Project administration (lead); 
Writing - original draft (equal); 
Writing - review \& editing (equal).

\subsection*{Data availability}
The data that support the findings of this study are available from the corresponding author upon reasonable request.

\appendix

\section{Derivation of the quantum state of light in Eq.~(1)}
\label{sec:appendix1}
\renewcommand{\theequation}{\ref{sec:appendix1}\arabic{equation}}
\setcounter{equation}{0}

We derived the quantum state of the entangled photon pairs produced by the SPDC process, explicitly accounting for the central time of the pump laser. We considered electric fields inside a one-dimensional nonlinear crystal. The positive frequency component of the electric field operator can be expressed as
\begin{align}
    \E_j^+(z,t)
    =
    C \int d\omega
    \,
    \hat{a}_j(\omega)
    e^{i k_j(\omega)z -i\omega t},
\end{align}
where the slowly varying envelope approximation is adopted, and all constants are combined into a factor $C$. For simplicity, the pump laser was treated as a classical field,
\begin{align}
    E_\p(z,t)
    =
    \int d
    \omega\,
    \alpha_\p(\omega)
    e^{i k_\p(\omega)z -i\omega (t-t_\p)},
\end{align}
where $\alpha_\p(\omega)$ denotes the normalized pump envelope function, and $t_\p$ is the central time of the pump laser. The Hamiltonian governing the SPDC process can be written as \cite{Christ:2011ku}
\begin{align}
    \hat{H}_\mathrm{PDC}(t)
    =
    \int_{-L/2}^{L/2} dz 
    \,
    \chi^{(2)}
    E_\p(z,t)
    \E_\s^-(z,t)
    \E_\i^-(z,t)
    +
    \mathrm{h.c.},
    \label{eq:PDC-Hamiltonian}
\end{align}
where $\chi^{(2)}$ is the second-order susceptibility of the crystal. By using the Hamiltonian in Eq.~\eqref{eq:PDC-Hamiltonian}, the unitary evolution of the state vector can be expressed as
\begin{align}
	\lvert \psi_\twin \rangle
    =
    \hat{\mathrm{T}}
    \exp
    \left[
    -\frac{i}{\hbar}
    \int_{-\infty}^{\infty} dt
    \,
    \hat{H}_\mathrm{PDC}(t)    
    \right]
	\lvert \vac \rangle,
    \label{eq:psi-PDC}
\end{align}
where $\hat{\mathrm{T}}$ denotes the time-ordering operator. The weak-down-conversion regime was analyzed; therefore, the time ordering was ignored. The vector state in Eq.~\eqref{eq:psi-PDC} can be approximated as
\begin{align}
	\lvert \psi_\twin \rangle
    &\simeq
    -\frac{i}{\hbar}
    \int_{-\infty}^{\infty} dt
    \,
    \hat{H}_\mathrm{PDC}(t)
    \vert \vac \rangle
\notag \\    
    &=
    \zeta
    \iint d\omega_1 d\omega_2
    \,
    e^{i(\omega_1+\omega_2)t_\p}
    \alpha_\p(\omega_1+\omega_2)
\notag \\
    &\quad\times
    \sinc\frac{\Delta k(\omega_1,\omega_2)L}{2}
	\adagger_\s(\omega_1) \adagger_\i(\omega_2)
	\vert \vac \rangle,
\end{align}
where $\Delta k(\omega_1,\omega_2) = k_\p(\omega_1+\omega_2) - k_\s(\omega_1) - k_\i(\omega_2)$ represents the wave vector mismatch, $\zeta$ gathers all constant factors and corresponds to the conversion efficiency of the SPDC process.


\section{Stimulated emission contribution to the two-dimensional electronic spectrum}
\label{sec:appendix4}
\renewcommand{\theequation}{\ref{sec:appendix4}\arabic{equation}}
\setcounter{equation}{0}
\label{appendixB}

The stimulated emission contribution to the 2D electronic spectrum can be written as \cite{Fujihashi:2015kz}
\begin{align}
    \mathcal{S}_\mathrm{SE}(\omega_3, \tau_2, \omega_1)
    =
    \mathcal{S}_\mathrm{SE}^{\rephasing}(\omega_3, \tau_2, \omega_1)
    +
    \mathcal{S}_\mathrm{SE}^{\nonrephasing}(\omega_3, \tau_2, \omega_1),
\end{align}
where
\begin{align}
    \mathcal{S}_\mathrm{SE}^{(x)}(\omega_3,\tau_2,\omega_1)
    =
    \int d\tau_3 e^{i\omega_3 \tau_3}
    \int d\tau_1 e^{\mp i\omega_1 \tau_1}
    \Phi_\mathrm{SE}^{(x)}(\tau_3,\tau_2,\tau_1)  
\end{align}
with the upper and lower signs in $\mp$ corresponding to the rephasing and nonrephasing contributions, respectively.


%


\end{document}